\documentclass[12pt]{article}
\usepackage{epsfig}
\input amssym.def
\input amssym.tex

\def \ol{\overline}
\def\be{\begin{equation}}       
\def\ee{\end{equation}}         

\def\bea{\begin{eqnarray}}      
\def\eea{\end{eqnarray}}
                                
\def\ba{\begin{array}}
\def\ea{\end{array}}
\def\bd{\begin{displaymath}}
\def\ed{\end{displaymath}}

\def\eq{\begin{equation}}
\def\eqe{\end{equation}}
\def\eqa{\begin{eqnarray}}
\def\eqae{\end{eqnarray}}
\def\ena{\end{eqnarray}} 
\def \ol{\overline}

\def\eg{{\it e.g.~}}

\def\nn{\nonumber}
\def\Tr{{\rm Tr}}
\def\tr{{\rm tr}}

\def\unit{1 \hskip-.3em \raise2pt\hbox{$ \scriptstyle |$ } }

\def\a{\alpha}
\def\b{\beta}
 
\def\d{\delta}
   
\def\g{\gamma}

\def\k{\kappa}                    
\def\l{\lambda}
\def\m{\mu}
\def\n{\nu}
  
\def\r{\rho}                                     
\def\s{\sigma}                                   

\def\F{\Phi}

\def\cn{{\cal N}}
\def\co{{\cal O}}

\def\bd{\begin{displaymath}}
\def\ed{\end{displaymath}}
\def\quart{\frac14}
\def\6{\partial}
\def\N4{{\cal N}=4}
\def\lab{\label}
\def\bq{\bar{q}}

\def\half{{1 \over 2}}

\def\bop#1{\setbox0=\hbox{$#1M$}\mkern1.5mu
        \vbox{\hrule height0pt depth.04\ht0
        \hbox{\vrule width.04\ht0 height.9\ht0 \kern.9\ht0
        \vrule width.04\ht0}\hrule height.04\ht0}\mkern1.5mu}

\def\>{\rangle} 

\def\<{\langle} 
\def\Dsl{D \hskip-.6em \raise1pt\hbox{$ / $ } }
\def\to{\rightarrow}

\def\+{\oplus}

\def\half{{1 \over 2}}
\def\Tr{{\rm Tr}\, }

\begin{document}
\begin{flushright}
MIT-CTP-3360 \\
{\tt hep-th/0506128}
\end{flushright}

\begin{center}
{\Large\bf
Comments on the $\b$-deformed $\N4$ SYM Theory}
\vskip .6truecm

{\large\bf Daniel Z. Freedman${}^{\star\dagger}${}\footnote{
{\tt dzf@math.mit.edu}}\\ 
 and Umut G\"ursoy${}^{\dagger}${}\footnote{
{\tt umut@mit.edu}}} \\
\vskip 0.3truecm
 { *\it Department of Mathematics\\ 
*${}^{\dagger}$  Center for Theoretical Physics\\
 Massachusetts Institute of Technology\\
Cambridge MA 02139, USA }
\vskip 0.3truemm
\end{center}
\vskip .3truecm
\begin{abstract}

Several calculations of 2- and 3-point correlation functions
in the deformed theory are presented. The
central charge in the Lunin-Maldacena gravity dual is shown
to be independent of the deformation parameter. Calculations
show that 2- and 3-point functions of chiral primary 
operators have no radiative corrections to lowest order in
the interactions. Correlators of the operator $\tr(Z_1Z_2)$, 
which has not
previously been identified as chiral primary, also have 
vanishing lowest order corrections. 

\end{abstract}

\newpage

\section{Introduction}

In the $\b$-deformed $\N4$ SYM theory, the 3 $\cn =1$ 
chiral superfields $\F_i$ have interaction superpotential
\be \label{w}
W = h \Tr(q \F_1\F_2\F_3 - \frac{1}{q} \F_1\F_3\F_2), 
\ee
with complex parameters $h$ and $q$, the latter usually 
written as $q= e^{i \pi \b}$. The undeformed
$\cn =4$ theory is obtained for $q=\pm 1$ and $|h| = 2g$,
where $g$ is the gauge coupling. The deformed theory has $\cn=1$
SUSY with $U(1)_R$ and $U(1)\times U(1)$ flavor symmetries.
It was pointed out in \cite{leighst} that 
renormalization group beta functions vanish if one
condition on the constants $q,h$ is satisfied, giving 
an exactly marginal deformation of $\cn =4$ SYM with $\cn =1$ 
SUSY.  
New attention has been focused on this deformed theory,
for gauge group $SU(N)$, after its gravity dual was found by 
Lunin and Maldacena\footnote{
Earlier partial results were obtained in \cite{0205090}}
\cite{lm}. It is a solution of Type IIB supergravity with
product metric $AdS_5\times \tilde{S}_5$, where $\tilde{S}_5$ is
a deformed 5-sphere. The solution has many interesting features.
Further
developments have appeared in the subsequent recent literature 
\cite{0502083}-\cite{cml7}.

In this note, we present several results most of which concern 
the
properties of 2- and 3-point functions of $SU(2,2|1)$ chiral
primary operators in the deformed theory in the weak coupling
limit, thus exploring the perturbative dynamics of the exactly
marginal deformation. 
For this purpose we combine the methods of \cite{dhfs} and
Appendix D of \cite{g7}. The major conclusion is that the order
$\l = g^2N$ radiative corrections to these correlators vanish,
which is evidence that they are ``protected'', that is given
exactly by their free field values. In the undeformed $\cn =4$ 
theory, this ``protected'' property was first observed in the
strong coupling limit through the $AdS_5\times S_5$ gravity dual
\cite{lmrs}. It may be possible to carry out a similar 
analysis  for the Lunin-Maldacena solution  on the deformed
$\tilde{S}_5$,   although the
problem of Kaluza-Klein decomposition of the coupled fluctuations
is quite difficult.  The spin chain method of
calculation of anomalous dimensions, initiated in \cite{mz} for
the $\cn =4$ theory, has also been applied to the 
$\b$-deformation
in \cite{frt}, and information on 3-point functions may
also be available through this method \cite{oku}. Spin chain
techniques are elegant and useful, but the simple weak coupling
methods used herein lead to a clear and worthwhile picture of 
the correlators of chiral primaries.

\section{The Central Charge from the Gravity Dual}

If the Lunin-Maldacena solution is indeed the dual of the
$\b$-deformed $\cn =4$ theory, then correlation functions
computed from gravity by techniques developed early in the
study of the AdS/CFT correspondence \cite{gkp,w,mit1,mit2}
should agree with field theory correlators in the limit
$N \to \infty$ and $\l = g^2N >> 1$. Operators in the gauge
theory are dual to fluctuations of IIB gravity theory fields 
about the L-M background solution. In most cases therefore,
analysis of the fluctuations is required before correlation
functions can be studied, and this is difficult as stated above.

One important case in which detailed fluctuation analysis is
not required is the 2-point function $\<T_{\m\n}T_{\r\s}\>$ 
which determines the central charge $c$ of the CFT. An exactly
marginal deformation should not change the central charge from
its value $c= N^2/4$ in the undeformed 
$\cn = 4$ theory. The gravity calculation of this correlator
requires only the metric fluctuations in the $AdS_5$ part of the
metric. We are thus interested in reducing the gravitational part
of the IIB action on the $\tilde{S}_5$ internal space, so we
write
\bea  \lab{gract}
S_{10} &=&\frac{1}{2\k_{10}^2} \int d^{10}x \sqrt{G_{10}}
\left[R_{10} + \ldots \right]\\
      &=&  \frac{A}{2\k_{10}^2}\int d^5x \sqrt{g_{5}}\left[R_{5} +12/L^2 + 
\ldots\right ].
\eea
For a direct product metric, the factor $A$ 
would be $Vol(\tilde{S}_5)$, as first shown in 
\cite{hensken} and further applied in \cite{gub}. Things
are slightly more complicated in the L-M case because the 
Einstein frame metric is a warped product given by
\be \lab{metric}
ds_E^2 = L^2 G^{-1/4}[ ds^2_{AdS_5} + \sum_i(d\mu_i^2 + G\m_i^2
d\phi_i^2) + |\tilde{\g}|^2 L^4 G  \m_1^2\m_2^2\m_3^2
(\sum_i d\phi_i)^2],\\
\ee \bea
G^{-1} &=& 1 + |\tilde{\g}|^2
L^4(\m_1^2\m_2^2+\m_2^2\m_3^2+\m_1^2\m_3^2),\\
L^4 &=& 4\pi N.
\eea
The $\tilde{S}_5$ metric is parameterized by 3 angles $\phi_i$
and 3 positive coordinates $\m_i$ which are constrained to
satisfy $\sum_i \m_i^2 = 1$. The complex parameter $\tilde{\g}$
is the parameter in the L-M solution which describes a general
$\b$-deformation with $\b = \g + i\s$. We are using (3.24) of 
\cite{lm}. Denoting 
the $\tilde{S}_5$ metric in (\ref{metric}) by 
$ds^2 = h_{ij} dy^idy^j = \sum_i (d\m_i^2 + \ldots)$, we see
that the factor $A$ in (\ref{gract}) is given by
\be \lab{meas}
A = \int dy^1\ldots dy^6 \sqrt{h} G^{-5/4}G^{1/4} \delta( \sum_i \m_i^2
- 1).
\ee
It is now straightforward to calculate 
$h= \det{h_{ij}} = G^2 \m_1^2\m_2^2\m_3^2$ and to
observe that the {\it integrand} in (\ref{meas}) is independent of
the deformation parameter $\tilde{\g}$. This is sufficient to
show that the deformation, as described by the L-M solution, does 
not change the central charge $c$. The central charge in the
 L-M solution was also studied in \cite{pal}. 

Actually there is a general argument that $n$-point functions of $T_{\m\n}$ are
unchanged by the deformation. Recall that the L-M solution is
generated by an $SL(2,R)$ transformation which changes the
modular parameter of a 2-torus in the 10 dimensional 
spacetime, but leaves
the Einstein frame metric of the complementary 8 dimensions
invariant. A $U(1)\times U(1)$ group, dual to the flavor group in the field theory, acts on the torus.
Bulk fields which are invariant under $U(1)\times U(1)$, 
such as
the 5-dimensional metric dual to the stress tensor, 
give boundary correlators which are unaffected by the
$SL(2,R)$ transformation which produces the deformation.
\footnote{We are grateful to Juan Maldacena who supplied this 
argument.}

\section{Two-point correlation functions at weak coupling.}

The main purpose of this section is to study the lowest
order radiative corrections to 2-point functions of various
scalar operators in the deformed theory. Among other things,
we will confirm the absence
of anomalous dimensions for the chiral primary operators 
identified in \cite{bl,bjl,lm}, and we will find that an
operator not in this list has vanishing anomalous dimension
in lowest order. The techniques we develop here will be applied
to 3-point correlators in the next section. It is important that
we use the gauge group SU(N), since the deformed U(N) theory is 
not conformal.

\subsection{Preliminaries} 

In a general $\cn = 1$ SUSY gauge theory with chiral superfields
$\Phi^A$ and general cubic superpotential
 \be\label{spot}
W = \frac16 Y_{ABC}\F^A\F^B\F^C,
\ee
the $\b$-function for the couplings $Y_{ABC}$ is determined by
the anomalous dimension matrix of the elementary $\F^A$ via
\be
\beta_{ABC} = Y_{ABD}\g^D_C +Y_{ADC}\g^D_B + Y_{DBC}\g^D_A.  
\ee
To 1-loop order the anomalous dimension matrix is \cite{jjn}
\be\label{gam1}
16\pi^2\g^A_B = \half Y^{BCD}\bar{Y}_{ACD}-2g^2C(R)^A_B,\qquad
C(R)^A_B = (T^aT^a)^A_B,
\ee
It may also be seen from (6b) of \cite{jjn} that if $\g^{(1)} =0$
and \newline
$Q=T(R) - 3C(G) =0$, then $\g^{(2)}$ also vanishes. Then the
gauge coupling $\b$-function $\b_{NSVZ}$ also vanishes 
through 3-loop order. 

In the $\b$-deformed $\cn = 4$ theory,  $\F^A\to \F^a_i$, 
while 
$C(R) = N \d^{ab}\d_{ij}$, \newline $C(G)=N$, and $T(R) = 3N$, so $Q=0$.
Color symmetry and the $Z_3 \otimes Z_3$ symmetry, see
\cite{0205090},  of the superpotential (\ref{w}) require
that $\g$ is diagonal, i.e. $\g^A_B\to\g \d^{ab}\d_{ij}$.
One readily obtains from (\ref{gam1}) 
\be\label{adim1}
\g^{(1)} = \left\{\frac14
  |h|^2\left(N(|q|^2+\frac{1}{|q|^2})-\frac2N
    |q -\frac1q|^2)\right)
-2g^2 N\right\}.
\ee
The condition $ \g^{(1)} =0$ ensures that the deformed $SU(N)$
theory is conformal invariant through 2-loop order. This
condition may be modified by higher loop corrections, but the
diagonal property ensures that there is always a fixed point set
of real codimension 1 in the space of couplings $h,q,g^2.$
The fixed line of $\cn =4$ SYM theory is attained when $q= \pm 1$
and $|h|^2 = 4 g^2$.

For gauge group $U(N)$, the situation is somewhat different.
The color singlet fields $\F^0_i$ have no D-type
interactions, but they do couple through the deformed 
superpotential
\be \lab{defw}
W_{U(N)} = \frac{1}{\sqrt{N}} h \F^0_1 (q-\frac1q)\tr(\F_2\F_3)
+ cyclic + W_{SU(N)}.
\ee
Singlet and non-singlet fields have different anomalous
dimensions, namely
\bea
16\pi^2 \g_{sing} &=& \frac{N}{4} |h|^2 |q-\frac1q|^2 \\
16\pi^2 \g_{non-sing} &=& \frac{N}{4} |h|^2 (|q|^2 +\frac{1}{|q|^2})
- 2 g^2N.
\eea
The marginal deformation is lost, but singlet
couplings flow to zero at long distance. The IR theory is
conformal with the same Lagrangian description as the $SU(N)$
theory.

We are interested in the 2-point (and  later 3-point) 
correlators of single trace
operators of the general form tr$(Z_1^jZ_2^kZ_3^l)$ where $Z_i$ is
the lowest component of the superfield $\F_i$. We now make
two observations which greatly simplify the required
calculations. We refer to the component Lagrangian which
is given in the Appendix, see (\ref{lagr}). The first observation 
concerns the relation of self-energy
corrections to internal scalar lines in the undeformed and
deformed theories. These corrections include a spinor loop with
F-term vertices from the Yukawa interaction in (\ref{lagr}).
 The amplitude is proportional to the term involving
$|h|^2$ in $\g^{(1)}$ in (\ref{adim1}) and thus appears to
depend on the deformation parameter $q$. However we must work
on the fixed point locus where $\g^{(1)}=0$, so the spinor loop
is really proportional to $g^2$ and independent of $q$. 

This fact
allows us to apply the arguments of \cite{dhfs} and Appendix B of 
\cite{g7} that order $g^2N$ radiative corrections due to 
D-term interactions cancel with self-energy insertions in 2- and 3-point
functions. The
reason is that D-term effects do not depend on the flavor 
quantum number of the fields $Z_i$. So their contribution
to correlators of tr$Z_1^jZ_2^kZ_3^l$ is the same as for the
the single flavor operators tr$Z_1^{j+k+l}$. The lowest order
radiative corrections to these correlators, which have no
F-term contributions (other than self-energy) then vanish by 
the combinatoric arguments of \cite{dhfs}. These arguments
are valid to lowest order in $g^2N$ and all orders in $1/N$.
They imply that we need only consider contributions from the
quartic F-term interaction in (\ref{lagr}).

Since we will be primarily interested in chiral primary
operators, we now describe the chiral ring of the $\b$-deformed
theory .  To classify these operators we define 
three separate $U(1)$ flavor groups such that the field $Z_i$ carries
unit charge under the $i$th $U(1)$ and the other fields  are uncharged.  
For general $\b$ the single trace operators of the chiral ring 
have the following charge assignments, 
\be\label{cr1}
(J,0,0),\quad (0,J,0),\quad (0,0,J),\quad (J,J,J).
\ee
Special chiral operators (dual to strings which wind contractible
cycles in $\tilde{S}_5$ \cite{lm})  exist for the special values
$\b=m/n$ where $m$ and $n$ are mutually prime integers. They
carry the $U(1)$ charges
\be\label{cr2}
(n_1,n_2,n_3), \qquad n_1=n_2=n_3 \;\;({\rm mod} ~n).  
\ee
As we will see from explicit calculations in the case $(J,J,J)$,
 the actual operator with vanishing anomalous
dimension is not tr$(Z_1^JZ_2^JZ_3^J)$ but rather a specific sum
over permutations of the $Z_i$ fields involved.
The same is true for the special cases $(n_1,n_2,n_3)$. 


\subsection{Correlators of  $\tr(Z_i^J)=\co^J_i.$}

Applying the arguments above, it is obvious that the 2-point
functions $\<\co_i^J~\bar{\co_i^J}\>$ (no sum on $i$) have no lowest order
radiative corrections, since the quartic F-term interaction does
not contribute.  In fact, to  lowest order,  {\it all}  correlators of  $\co_i^J$
agree with those in the undeformed theory. Thus all 3-point
functions  and all extremal $n$-point functions \cite{extr},  i.e.
\be \lab{ext}
\<  \co_i^{J_1}~\ldots \co_i^{J_k}~\bar{\co}_i^{J_1+\dots+J_k}\>
\ee
are protected at lowest order. However, non-extremal correctors
such as
\be \lab{nonex}
\<  \co_i^{J_1}\co_i^{J_2}~\bar{\co}_i^{J_3}\bar{\co}_i^{J_4}\>
\ee
are not protected, since they have non-vanishing $D$-term radiative
corrections. 

\subsection{The operators tr$(Z_i^JZ_j)$ for $i \ne j$ and 
$J>1$}

In the $\cn = 4$ theory these operators have protected 2- and
3-point correlators because they are obtained by applying an
SU(3) lowering operator to tr$Z_i^{J+1}$ which is clearly part
of the symmetrized traceless tr$X^k$ in the common $SO(6)$
designation. Since the SU(3)
symmetry is broken to $U(1)\otimes U(1)$ by the deformation we
would not expect that these operators remain chiral primary.

We now develop the ``effective operator'' method, see 
Appendix
D of \cite{g7}, which we use in most of our calculations.
This method allows us to obtain full results for two-point
functions with only half the combinatoric labor.
 
To be definite we take $i=1, j=2$. Other cases can be trivially
obtained from this one. Radiative corrections to the 2-point 
correlator
$\<\tr Z_1^JZ_2(x) \tr \bar{Z}_2\bar{Z}_1^J(y)\>$ are obtained
in two stages. In
the first step we calculate the effective operator obtained
from the Wick contractions of $Z(x)$ and $\bar{Z}(z)$ fields in
one factor of the F-term vertex in (\ref{lagr}),
\be 
\tr(Z_1^JZ_2) \tr\left(T^a (\bar{q}\bar{Z}_2\bar{Z}_1 -
\frac{1}{\bar{q}} \bar{Z}_1\bar{Z}_2)\right).
\ee
We ignore propagator factors temporarily, but keep track
of the $q$ dependence. To simplify the calculations we 
keep only leading terms in $N$ from the splitting/joining 
rules (\ref{join})
and we drop Wick contractions corresponding to nonplanar
diagrams. It is quite straightforward to obtain the 
effective operator
\be
\co = (\bar{q}-\frac{1}{\bar{q}})N \tr(Z_1^{J-1} a) 
\ee
In the second step we contract with the trace factor in
the conjugate operator, i.e.
\be
(Z_1^{J-1} a)(a \bar{Z}_1^{J-1}) = (Z_1^{J-1}\bar{Z}_1^{J-1})= N^J
\ee
Finally we put things together, starting with the free-field term
containing $J+1$ propagators and a factor $N^{Z+1}$ from
processing the traces. We then add the above result for the
interaction with propagators restored and the regulated 
integral
(\ref{bubb}) used. The correlator is then
\be \lab{2pt}
\<\tr Z_1^JZ_2(x) \tr \bar{Z}_2\bar{Z}_1^J(y)\>=
\frac{N^{J+1}}{(4\pi^2)^{J+1}} \frac{1}{(x-y)^{2(J+1)}}[ 1-\g\ln(M^2(x-y)^2)]
\ee
with anomalous dimension $\g = |h|^2N |q-\frac1q|^2/8\pi^2$.
Note that $\g$ vanishes when $q \to \pm 1$ which is the limit
of the undeformed $\cn =4$ theory.

The leading $N$ approximation is a major simplification for $J>1$
and it is appropriate for comparison with
gravity results using AdS/CFT. Of course it gives an incomplete
answer in the field theory. 

\subsection{The special case tr$(Z_iZ_j)$ for $i \ne j$}

In this case it is easy to include terms of all orders in 
$N$. 
The effective operator is
\bea
\co &=&\tr(Z_iZ_j) \tr\left(T^a (\bar{q}\bar{Z}_j\bar{Z}_i -
\frac{1}{\bar{q}} \bar{Z}_i\bar{Z}_j)\right)\\
&=& (\bar{q}-1/\bar{q})[\frac{N^2-1}{N}(1-\frac1N)(T^a)]
\eea
which vanishes in $SU(N)$, indicating that to lowest order 
the operator behaves as a chiral primary. 

Of course, this could be an
accident of lowest order perturbation theory, and higher loop
corrections should be studied. At 3-loop order there are
many contributing Feynman diagrams including both quartic and
Yukawa interactions from (\ref{lagr}), so this appears to 
be a difficult problem. Supergraph methods, as used in
\cite{penati}, may be advantageous.\footnote{In \cite{Zanon} it has
been shown that the anomalous dimension of tr$(Z_iZ_j)$ vanishes to
3-loop order.} 

It is curious to repeat the lowest order
calculation for gauge group $U(N)$, the effective
operator becomes
\be
\co = (q-\frac1q)N^{\frac32}\d^{a0}.
\ee
The 2-point function then takes the form (\ref{2pt}) with 
$\g = |h|^2N|q-\frac1q|^2/8\pi^2$ and $J=1$. So the operator
acquires anomalous dimension in this case.

\subsection{$(1,1,1)$ operator}

In this sector we expect that there is one linear combination
$\tr(Z_1Z_2Z_3) +\a\, \tr(Z_1Z_3Z_2)$ which has vanishing
anomalous dimension. We apply the splitting joining rules
in (\ref{sun}) to work out the effective operators
\bea\lab{efopp}
\co_{123} =\tr(Z_1Z_2Z_3)\tr\left(T^a (\bar{q}\bar{Z}_2\bar{Z}_1 -
\frac{1}{\bar{q}} \bar{Z}_1\bar{Z}_2)\right)\nn\\
=[N\bar{q}-\frac{2}{N}(\bar{q}-\frac{1}{\bar{q}})]\tr(Z_3T^a)\nn\\
\co_{132} =\tr(Z_1Z_3Z_2)
\tr\left(T^a (\bar{q}\bar{Z}_2\bar{Z}_1 -
\frac{1}{\bar{q}} \bar{Z}_1\bar{Z}_2)\right)\nn\\
=-[\frac{N}{\bar{q}}-\frac{2}{N}(\bar{q}-\frac{1}{\bar{q}})]\tr(Z_3T^a)\nn\\
\eea
One must add the similar contributions from the other two cyclic 
permutations in the $F$-term Lagrangian. These give contributions which differ 
from (\ref{efopp}) only by the replacements $\tr(Z_3 T^a)\to \tr(Z_1 T^a)$ and 
$\tr(Z_3 T^a)\to \tr(Z_2 T^a)$. 

The 1-loop anomalous dimension is obtained by contracting the effective
operators for the linear combinations $\tr(Z_1Z_2Z_3) +\a\, \tr(Z_1Z_3Z_2)$
with their conjugates. It is clear that the anomalous dimension vanishes if and
only if the operator itself vanishes. This  fixes 
the complex coefficient $\a$.  For arbitrary
  $q$ and $N$ one obtains the following BPS operator:
\be\label{BPS2}
\co^{(1,1,1)} = \tr(Z_1Z_2Z_3) +
  \frac{(N^2-2)\bar{q}^2+2}{N^2-2+2\bar{q}^2}
\,\,\tr(Z_1Z_3Z_2).
\ee
To leading order in $N$, the coefficient $\a \to \bar{q}^2$.

\subsection{$(2,2,0)$ operator}

It is straightforward to use the effective operator method to investigate
the linear combination  $\tr(Z_1Z_1Z_2Z_2)  +c\,\tr(Z_1Z_2Z_1Z_2)$.
To leading order in $N$, the effective operator is
\be \lab{oeff2}
\co = N[(2c\bar{q}-\frac{1}{\bar{q}})\tr(Z_1Z_2T^a) +(\bar{q} -\frac{2c}{\bar{q}})\tr(Z_2Z_1T^a)].
\ee
The anomalous dimension vanishes  to 1-loop order only if $q^4=1$, which
is what we expect from the classification of chiral primary operators.
There are two cases to consider:\newline
i) $q=\pm 1$ with $ c=\frac12$.  This means that we have 
the undeformed
$\cn =4 $ SYM theory, and the operator is the second $SU(3)$ descendent 
of the primary $\tr(Z_1^4)$.\newline
ii) $q=\pm i$ with $c=-\frac12$. This is the expected chiral primary chiral
operator of type $(2,2,0)$ which occurs for the special value $\b=1/2$.
With  $1/N$ corrections included, this operator becomes
\be\label{BPS3}
\co^{(2,2,0)} = \tr(Z_1Z_1Z_2Z_2) -\half\left(\frac{N^2-8}{N^2-4}\right) \,\, 
\tr(Z_1Z_2Z_1Z_2). 
\ee

\subsection{BPS operators at large $N$}

So far, we constructed some special members of the chiral primary
operators. Our results above are valid for all values of $N$. It
becomes cumbersome determine the chiral primaries that are composed
of more canonical fields by the method of
vanishing effective operator. However, one can easily work out the
general form of these operators when $N$ is large, as below. 

The $(J_1,J_2,J_3)$ operator is given as, 
\be\lab{cpop}
\co = \sum_{\pi} c_{\pi}\,\, \tr(\pi\cdot Z_1^{J_1}Z_2^{J_2}Z_3^{J_3})
\ee
where $\pi$ is the sum over all distinct permutations 
modulo cyclicity of the trace\footnote{It may be 
interesting to derive this definition of the chiral primaries 
from a $q-algebra$ method as in \cite{u}.}. Determination of the   
the complex coefficients $c_{\pi}$, turns out to be easy in the large $N$ limit:
\be\lab{cpi}
c_{\pi} = \frac{\bq^{2k_{\pi}}}{s_{\pi}},
\ee
where $k$ is a positive integer and $s$ 
is a symmetry factor of the permutation. 

To specify $k_{\pi}$ and $s_{\pi}$ we express the permutation $\pi$ in terms of the elementary 
exchanges (12), (23) and (31). We assign $c_1=1$ for the identity permutation. 
Then $k_{\pi}$ is obtained by introducing a factor $\bq^2$ for each exchange. 
For example, 
$$ 223311 \to \bq^{4} 212313.$$
The symmetry factor is the number of repeated arrays in the permutation: 
\eg for 213213 $s=2$, for 212121 $s=3$. Division by the symmetry factor is required for the 
cancellation among contributions to the effective operators.

Let us illustrate the basic mechanism for the cancelation of the one-loop
radiative corrections to (\ref{cpop}) with the special $(n,n,0)$
operators. They are chiral primary when  $\beta=m/n$. These 
operators are
conjectured to be dual
to strings that rotate along the
two contractible cycles in the deformed $S^5$ with momenta $n_1=n_2=n$
and winding numbers $w_1 = -m$, $w_2 = m$ \cite{lm}. 
Consider the F-term
interactions of the following array of fields.
\be\lab{exp}
(Z_2\cdots Z_2Z_1\cdots Z_1)
\ee
The quartic interaction that involves the pair of fields in the middle of
this array gives a contribution to the effective operator that is
proportional to
$$-\frac{1}{\bar{q}} (Z_2\cdots T^a \cdots Z_1)$$
in the limit $N\to\infty$. In case of the special BPS operator this
contribution cancels out the contribution that comes from a term in
(\ref{cpop}) that is obtained from (\ref{exp}) by shifting the
position of $Z_2$ one step right. This contribution is
proportional to,  
$$\bar{q}\left(\frac{1}{\bq}\right)^2 (Z_2\cdots T^a  \cdots Z_1)$$
in the limit $N\to\infty$ so it cancels out the first contribution above.  
Similarly, one can see that all of the F-term contributions are paired up in a
manner similarly and cancel out provided that $q^{2n} =1$. 

\begin{figure}[htb]
\centerline{\epsfxsize=15cm\epsfysize=6cm\epsffile{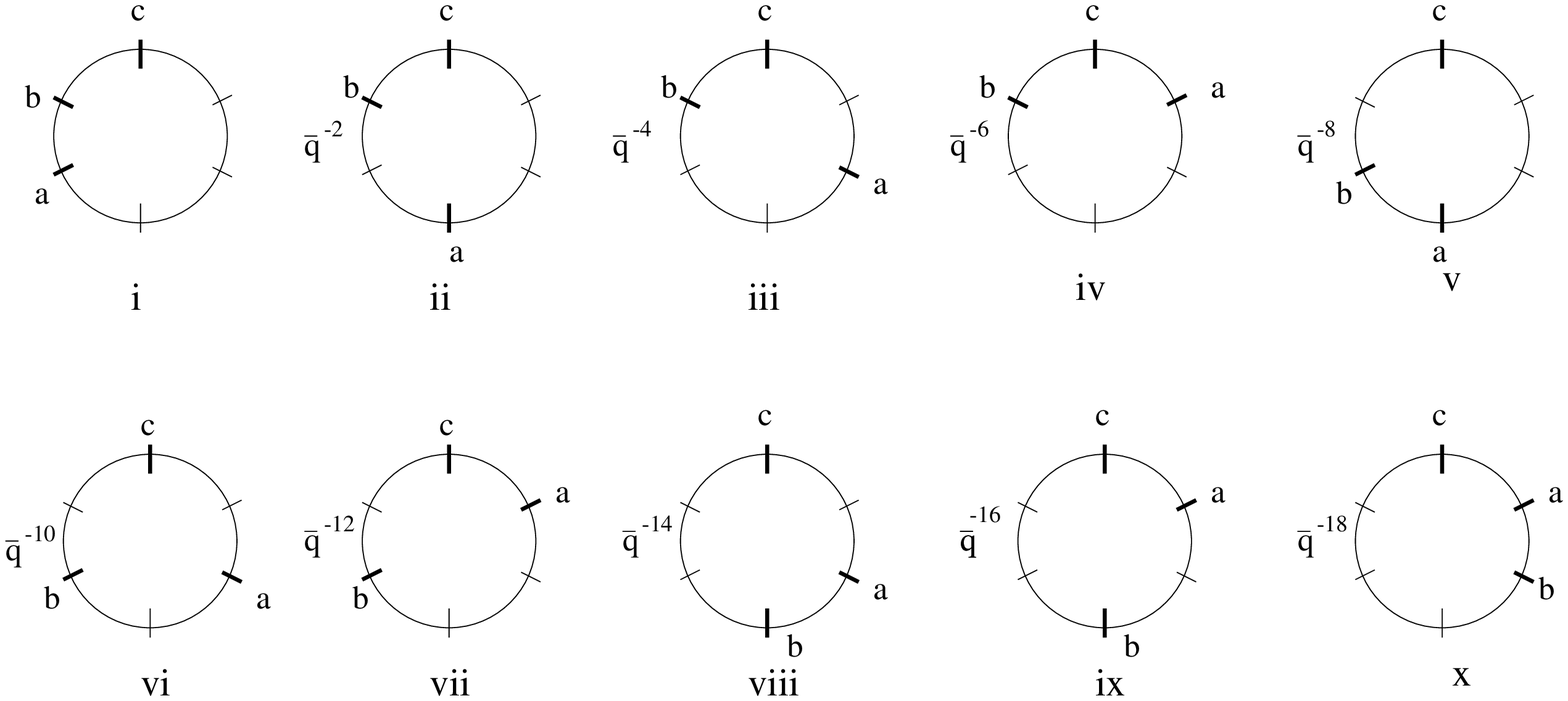}}
\caption{Terms in the $(3,3,0)$ operator shown together with their coefficients. 
It is easy to see that this expansion is the same as 
(\ref{spopn3}) when the condition $q^6=1$ is used. Bold marks denote the positions of 
$Z_2$ fields, and the smaller marks denote $Z_1$'s. 
F-terms interactions of $a$ with its neighbors 
is cancelled out among the group of terms  i-ii-iii-iv,  v-vi-vii 
and viii-ix seperately. Similarly the interactions of $b$ is cancelled out among the groups 
ii-v, iii-vi-viii and iv-vii-ix-x. The interactions of $c$ is cancelled by the 
same mechanism among the groups i-v-viii-x, ii-vi-ix and iii-vii {\em provided that} 
$q^6=1$.}   
\end{figure}

This last requirement can be seen as follows. There are $n$ $Z_2$ 
fields in the $(n,n,0)$ operator that pair up with their neighbor $Z_1$ fields to 
form an F-quartic vertex. Let us label these fields by $a$, $b$, $c,\dots$ 
as in the fig. 1. As can easily be seen from the figure the interactions of the 
first $n-1$ $Z_2$ fields cancel out by the above mechanism. However in order 
to cancel the interactions of the $n$th field by the same mechanism 
one needs to impose the cyclicity condition $q^{2n}=1$. 
Note that this requirement coincides with the definition of the chiral ring.  

We list several examples of the prescription (\ref{exp}). 
The case of $n=2$ is already discussed 
at the end of the previous section where it was simple enough to determine the operator 
for all orders in $N$.     
In the next case $n=3$ one has the following two special operators
(for large $N$),
\bea\lab{spopn3}
 \co^{(3,3,0)} &=& \tr(Z_2Z_2Z_2Z_1Z_1Z_1) + q^{2}\,\, 
\tr(Z_2Z_2Z_1Z_2Z_1Z_1) +\nn\\ 
{}&&+q^{4}\,\, \tr(Z_2Z_2Z_1Z_1Z_2Z_1)+ 
 \frac13\tr(Z_2Z_1Z_2Z_1Z_2Z_1) 
\eea  
where $q = e^{2i\pi/3}$ and $q = e^{4i\pi/3}$. These two
operators are related to each other by the relabeling of the fields
$Z_1 \leftrightarrow Z_2$. 

Finally, for $n=4$, q-variation method results in the following special BPS operators
(for large $N$),
\bea\lab{spopn4}
 \co^{(4,4,0)} &=& \tr(Z_2Z_2Z_2Z_2Z_1Z_1Z_1Z_1) + q^{2} \,\,
\tr(Z_2Z_2Z_2Z_1Z_2Z_1Z_1Z_1) \nn\\
{}&&+ q^{4}\,\, 
\tr(Z_2Z_2Z_2Z_1Z_1Z_2Z_1Z_1) + q^{6}\,\, 
\tr(Z_2Z_2Z_2Z_1Z_1Z_1Z_2Z_1)\nn\\
{}&& +q^{4} \,\,\tr(Z_2Z_2Z_1Z_2Z_2Z_1Z_1Z_1) +q^{6}\,\, 
\tr(Z_2Z_2Z_1Z_2Z_1Z_2Z_1Z_1)\nn \\
{}&& + \tr(Z_2Z_2Z_1Z_2Z_1Z_1Z_2Z_1) + 
\half \tr(Z_2Z_2Z_1Z_1Z_2Z_2Z_1Z_1)\nn\\
{}&& + q^{6}\,\,\tr(Z_2Z_2Z_1Z_1Z_2Z_1Z_2Z_1) 
+ \frac{q^{4}}{4}\tr(Z_2Z_1Z_2Z_1Z_2Z_1Z_2Z_1)\nn
\eea  
where the independant choices for $q$ are $e^{i\pi/4}$ and
$e^{i\pi/2}$ and $e^{i3\pi/4}$. 

It is straightforward to check that the F-term radiative corrections
 to the two-point functions of these operators vanish as
 $N\to\infty$. However we expect to obtain modifications for finite $N$.  

It is straightforward but somewhat more complicated to show 
that the prescription (\ref{cpop}) determines an operator with
vanishing anomalous dimension for all values of $J_1,J_2,J_3$
which correspond to chiral primaries.


 
\section{Three point functions of chiral primary operators}

In this section we consider three-point correlators of the
chiral primary operators listed in (\ref{cr1}, \ref{cr2}) and
their complex conjugates. We note that all three-point 
functions are extremal and can be schematically written as
$\<\bar{\co}\co_1 \co_2\>$ where $\co_1$ and $\co_2$ contain
only $Z_{1,2,3}$ and $\bar{\co}$ contains only conjugate 
fields. 

As discussed in Sec 3.1 above, the combinatoric arguments of 
\cite{dhfs} imply that the $D$-term interactions cancel
provided that one works on the fixed point locus where
$\g^{(1)}=0$. 

We now argue that lowest order
$F$-term interactions also vanish because the effective 
operator method gives the same operator which occurred in
the computation of two-point functions, and that operator 
thus vanishes. 
The one-loop radiative correction is given by Wick contractions of 
the three point function with the F-term quartic vertex:
\be\lab{F3} 
\sum_{i,a}\<\ol{\co}\frac{\6 W}{\6
  Z_i^a}\frac{\6 \bar{W}}{\6
  \bar{Z}_i^a}\co_1^{J_1}(y)\co_2^{J_2}\>.
\ee
Wick contractions of $\ol{\co}\frac{\6 W}{\6 Z_i^a}$ give the same effective 
operator $\ol{\co}^a_i$ which occurs in the computation of 
the F-term corrections to the 2-point function of $\ol{\co}$. These corrections 
vanish if $\ol{\co}$ is chiral primary. 
Therefore 
the one-loop radiative correction to the three point function is absent. 

These arguments generalize immediately to extremal $n$-point
functions of all chiral primaries.

\section{Conclusions}
We have shown that lowest order calculations by the effective
operator method confirm the assignment of chiral 
primary operators of \cite{bjl} and suggest that their 
two- and
three-point correlation functions (and extremal correlators)
have vanishing radiative corrections. It would be 
interesting to
confirm this suggestion in further study and to ascertain 
the properties of $\tr(Z_iZ_j)$, with $i \ne j$, which we have
shown to be protected to lowest order although it was not
previously recognized as a chiral primary. There are far 
fewer
chiral primary operators in the $\b$-deformed theory than 
in its 
undeformed $\cn =4$ parent. But it is striking that their
correlators appear to enjoy the same properties that have 
been
established in the latter case. 

\section{Acknowledgments}
The authors thank Juan Maldacena for much encouraging and 
useful
correspondence during the course of this work. We also thank
Ken Intriligator and Carlos Nu{\~n}ez for useful discussions.

\section{Appendix} 
\subsection{Conventions}

We use the following conventions:
\be\label{conv}
\tr(T^aT^b)=\delta^{ab}, \qquad [T^a,T^b]=i\sqrt{2}f^{abc}T^c,
\ee
With these conventions, the action of the deformed theory can be written in the $\cn=1$
component notation as follows:
\bea\label{lagr}
{\cal L}  &=&   \tr\bigg[\quart F_{\mu \nu }^{~~2}+\half\bar{\lambda}
             {D}\!\!\!\!\slash\lambda
             +\overline{D_{\mu }Z^i }D_{\mu }Z^i +\half 
             \bar{\psi}^i{D}\!\!\!\!\slash\psi^i 
+ g (\bar{\lambda}\bar{Z}^i L\psi^i
  -\bar{\psi}^i R Z^i\lambda) \nonumber\\
 {}&& 
+\frac{1}{4} g^{2}[\bar{Z}^i, Z^i]^{2}\bigg]\nonumber\\ 
{}&& + \frac{h}{2}
\tr\left(q\bar{\psi}_2 L Z_3\psi_1-\frac1q\bar{\psi}_1 L Z_3\psi_2\right)
+ \frac{\bar{h}}{2}\tr\left(\bar{q}\bar{\psi}_1 R
             \bar{Z}_3\psi_2-\frac{1}{\bar{q}}\bar{\psi}_2 R
             \bar{Z}_3\psi_1\right) \nn\\ 
{}&& +|h|^2\tr\left(T^a(q Z_2 Z_3-\frac1qZ_3Z_2)\right)
\tr\left(T^a(\bar{q} \bar{Z}_3
             \bar{Z}_2-\frac{1}{\bar{q}}\bar{Z}_2\bar{Z}_3)\right) +
             {\rm cyclic}\nn\\
\eea
where $D_{\m}Z=\6_{\m}-i\frac{g}{\sqrt{2}}[A_{\m},Z]$ and similarly 
for the fermions. The first two lines contain D-type 
interactions, and the last two lines give the F-type 
interactions. Note that cylic permutations of the labels 123
must be added in the F-terms. 
It is the quartic F-type interaction of the scalars
that plays the major role in our computations. 
The splitting/joining rules could be used to simplify this term, but the
factored form is the most convenient for the ``effective
operator'' method used in most of our calculations.

The Lagrangian above is valid for Euclidean signature in which
we consider Wick contractions with $e^{-S_{int}}$. The scalar
propagator is $\<Z(x) \bar{Z(y)}\> = 1/4\pi^2(x-y)^2$. The 
``bubble graph'' with 2-quartic vertices leads to the
space-time integral
\be \lab{bubb}
\frac{1}{(2\pi)^4} \int \frac{d^4z}{z^4(z-x)^4}~=~
\frac{\ln{M^2x^2}}{8 \pi^2 x^4}.
\ee
This result is obtained by (partial) differential regularization,
see \cite{fjl}. 

Our methd of calculation is based on the
splitting/joining rules for traces \cite{cvit,g7}. We present 
these in a
compact notation in which the generators of the fundamental
of $SU(N)$ are replaced by their index values, i.e. $T^a \to
a,~~a=1,\ldots ,N^2-1$,
and the trace of an arbitrary $N\times N$ matrix is denoted by
$\Tr(M) \to (M)$. The following rules can then be used to
evaluate traces and products thereof which involve sums over
repeated indices:
\bea \lab{join}
\begin{array}{rclrcl}
(MaM'a) &=& (M)(M')-\frac1N (MM') \qquad&(ab) &=& \delta^{ab}\\
(Ma)(aM') &=& (MM')-\frac1N (M)(M')\qquad&(a) &=& 0\\
 aa &=& \frac{N^2-1}{N} I&(I) &=& N
\end{array} \lab{sun}
\eea
To treat $U(N)$ we add the generator $T^0 = I/\sqrt(N)$. The
relations in the first column of (\ref{sun}) are valid for $U(N)$
if the $\frac1N$ terms are dropped. The relations in the second
column are valid except for the change $(a) \to \sqrt{N}\d^{a0}.$


\end{document}